\documentclass[prd,aps,twocolumn,a4paper,floatfix,nofootinbib]{revtex4-2}

\usepackage[utf8]{inputenc}
\usepackage{graphicx,psfrag}
\usepackage{mathrsfs}
\usepackage{amsmath,amsfonts,amssymb}
\usepackage{multirow}
\usepackage{diagbox}
\usepackage{comment}
\usepackage[dvipsnames]{xcolor}
\usepackage{enumerate}
\usepackage{booktabs}
\usepackage[normalem]{ulem}
\usepackage{hyperref}
\usepackage{mathtools}
\usepackage{siunitx}
\usepackage{textgreek}
\usepackage{float}
\usepackage{adjustbox}
\hypersetup{
    colorlinks = true,
    linkcolor = {blue},
    citecolor = {blue},
    urlcolor = {blue},
    linkbordercolor = {white},
    }

\usepackage{color}
\definecolor{cyan}{rgb}{0,0.9,0.9}
\definecolor{orange}{rgb}{0.9,0.5,0}
\definecolor{magenta}{rgb}{1,0,1}
\definecolor{purple}{rgb}{0.8,0.4,0.8}
\definecolor{gray}{rgb}{0.8242,0.8242,0.8242}
\definecolor{green}{rgb}{0.,0.8,0.}

\def\bam{{\textsc{bam}}}
\def\sgrid{{\textsc{sgrid}}}

\usepackage[normalem]{ulem}

\begin{document}

\title{Simulating Binary Neutron Star Mergers with Finite-temperature Equations of State: The influences of the slope of the symmetry energy and artificial heating}

\author{Henrique \surname{Gieg}$^{1}$}
\author{Maximiliano \surname{Ujevic}$^{2}$}
\author{Armen \surname{Sedrakian}$^{3,4}$}
\author{Tim \surname{Dietrich}$^{1,5}$}

\affiliation{${}^1$ Institut f\"ur Physik und Astronomie, Universit\"at Potsdam, Haus 28, Karl-Liebknecht-Str. 24/25, 14476, Potsdam, Germany}
\affiliation{${}^2$ Centro de Ci$\hat{e}$ncias Naturais e Humanas, Universidade Federal do ABC, 09210-170, Santo Andr{\'e}, 9210-170, SP, Brazil}
\affiliation{${}^3$  
Institute of Theoretical Physics, University of Wroc\l{}aw, 50-204 Wroc\l{}aw, Poland}
\affiliation{${}^4$  
Frankfurt Institute for Advanced Studies, D-60438 Frankfurt am Main, Germany}
\affiliation{${}^5$ Max Planck Institute for Gravitational Physics (Albert Einstein Institute), Am M\"uhlenberg 1, Potsdam 14476, Germany}

\date{\today}

\begin{abstract}
We present a new set of numerical-relativity simulations of merging binary neutron stars, aiming to identify possible observable signatures of the slope of the symmetry energy $L_{\rm sym}$. To achieve this goal, we employ a set of equations of state based on a parameterization of the covariant density functional theory of nuclear matter that allows controlled variations of $L_{\rm sym}$ and the skewness $Q_{\rm sat}$, holding the latter fixed. For a set of our simulations, we identify a steep energy gradient in the equation of state at subsaturation densities, which acts as a source of heating with subsequent stiffening produced by thermal support. Accounting for related structural modifications in the tidal deformability reconciles our results with theoretical expectations. On the other hand, we show that gravitational waves are unlikely to distinguish the role of $L_{\rm sym}$.
In contrast to this, we find that the ejecta composition is significantly altered in our simulations, which employ an M1 moment scheme, when $L_{\rm sym}$ is varied.
Based on our extracted dynamical ejecta properties, we compute r-process yields and find that they are distinct for the different $L_{\rm sym}$, especially at lower mass numbers $A \lesssim 120$. This suggests that electromagnetic counterparts are more likely to exhibit signatures; however, a direct connection to $L_{\rm sym}$ remains a challenge, given the complex interplay between details of the ejecta properties and the kilonova signal.
\end{abstract}

\maketitle

\section{Introduction}
\label{sec:Intro}

The detection of gravitational waves (GWs) from the collision of two neutron stars (NSs) in 2017~\cite{LIGOScientific:2017vwq} showed the potential of GW astronomy to probe matter at supranuclear densities and to provide constraints on the equation of state (EOS) governing the interior of NSs~\cite{LIGOScientific:2018cki,LIGOScientific:2018hze}. In fact, combined with 
other astrophysical measurements of neutron stars and with the electromagnetic counterparts of BNS mergers, multi-messenger astronomy has become an important field of research contributing to known nuclear physics constraints from theoretical nuclear-physics computations and experiments; see e.g.~\cite{Koehn:2024set,Chatziioannou:2024tjq} for recent reviews. Although no multi-messenger detection of a binary neutron star (BNS) merger happened after 2017, there have been several possible BNS mergers either seen via GWs, e.g., GW190425~\cite{LIGOScientific:2020aai}, or through electromagnetic observations, e.g., GRB211211A~\cite{Troja:2022yya,Rastinejad:2022zbg} or GRB230307~\cite{ JWST:2023jqa}. This increasing number of possible BNS candidates makes it clear that with the planned upgrades and construction of new telescopes, including third-generation GW detectors such as the Einstein Telescope~\cite{Punturo:2010zz,ET:2019dnz,Branchesi:2023mws,Abac:2025saz} and Cosmic Explorer~\cite{Reitze:2019iox,Evans:2023euw}, we will be confronted with a large number of BNS detections, which will offer new opportunities to uncover the physics of BNSs.

To extract information from astrophysical measurements, one has to cross-correlate the observed data with theoretical models describing the observational signatures. This requires us to model the BNS coalescence, including the late inspiral, merger, and postmerger of the two stars. Because of the strong gravitational fields and the high velocities of the stars, this requires solving the field equations of general relativity coupled to the equations of general relativistic hydrodynamics. Shortly prior to the merger and in the post-merger phase, this can only be done with the help of numerical relativity (NR) simulations; cf.~\cite{Alcubierre:2008,Baumgarte:2010} for textbook introductions to the topic. 

In this article, we are following previous works done with the BAM code~\cite{Bruegmann:2006ulg, Thierfelder:2011yi}, e.g., \cite{Gieg:2022,Schianchi:2023uky}, and perform simulations employing a new set of finite-temperature nucleonic EOSs computed in Ref.~\cite{Tsiopelas:2024ksy}. The set of EOSs is based on the covariant density functional (CDF) theory of dense nuclear matter~\cite{Oertel:2016bki,Sedrakian:2023}. The underlying parameterization is chosen such that the slope of the symmetry energy $L_{\rm sym}$ of nuclear matter can be varied smoothly, keeping the remaining parameters that define the EOS nearly unchanged. In this article, we will use the publicly avaiable EOSs from the CompOSE database that are tabulated for three different values of the slope of the symmetry energy $L_{\rm sym}$, namely $30$, $50$, and $70$~MeV, and a fixed skewness of nuclear matter $Q_{\rm sat} = 400$~MeV~\cite{Tsiopelas:2024ksy}. Using EOSs from the same family allows us, in general, to understand (i) the extent to which BNS mergers will allow us to constrain a specific parameter (here $L_{\rm sym}$) and (ii) if previously found relations based on simulations using different families are consistent with our findings.  
Indeed, while the previous work employed a broad set of EOSs generated from more or less non-selective sample parameterizations, we deliberately focus on a smaller, well-structured set of EOS models—specifically, DDLS30, DDLS50, and DDLS70~\cite{Tsiopelas:2024ksy}. This targeted selection enables a clearer and more transparent analysis of the impact of the symmetry energy slope $L_{\rm sym}$ on the properties of neutron stars and BNS mergers. By isolating the effects of $L_{\rm sym}$, our approach allows for a more systematic understanding of its role in determining macroscopic static observables, such as radii, tidal deformabilities, and dynamically generated signatures such as GW amplitudes.  It is worth noting that the EOS models used here were previously utilized in Ref.~\cite{Kochankovski:2025} to investigate the physics of BNS mergers using a different set of numerical schemes and computational tools compared to those used in the present work. Furthermore, the primary focus in Ref.~\cite{Kochankovski:2025} was on examining the effects of hyperons on BNS merger dynamics, rather than exploring the impact of variations of the nuclear matter parameters, such as the symmetry energy slope $L_{\rm sym}$  and the skewness $Q_{\rm sat}$.  Consequently, the role of these nuclear parameters in shaping the macroscopic properties of neutron stars and the emitted GW signatures during mergers remains largely unaddressed in that context, motivating the systematic analysis carried out in this study.

This paper is structured as follows. In Sec.~\ref{sec:NuclearPhysics}, we give an overview of the nuclear physics input and the resulting initial data, such as the masses and tidal deformabilities of cold NSs. In Sec.~\ref {sec:Methods}, we describe the BNS setups including the initial data, artificial heating, and dynamical evolution. The results of numerical simulations are collected in Sec.~\ref {sec:Results}, where we discuss the merger, post-merger dynamics, and the GW signal, and elemental abundances. Our conclusions are given in Sec.~\ref{sec:Conclusions}. 
In this article, we apply a metric with $(-,+,+,+)$ signature and geometric units with $G=c=M_\odot=1$, unless otherwise specified. 

\section{Nuclear Physics Input}
\label{sec:NuclearPhysics}
\subsection{Cold Equations of State}

\begin{figure*}[htpb!]
    \centering
    \includegraphics[width=\linewidth]{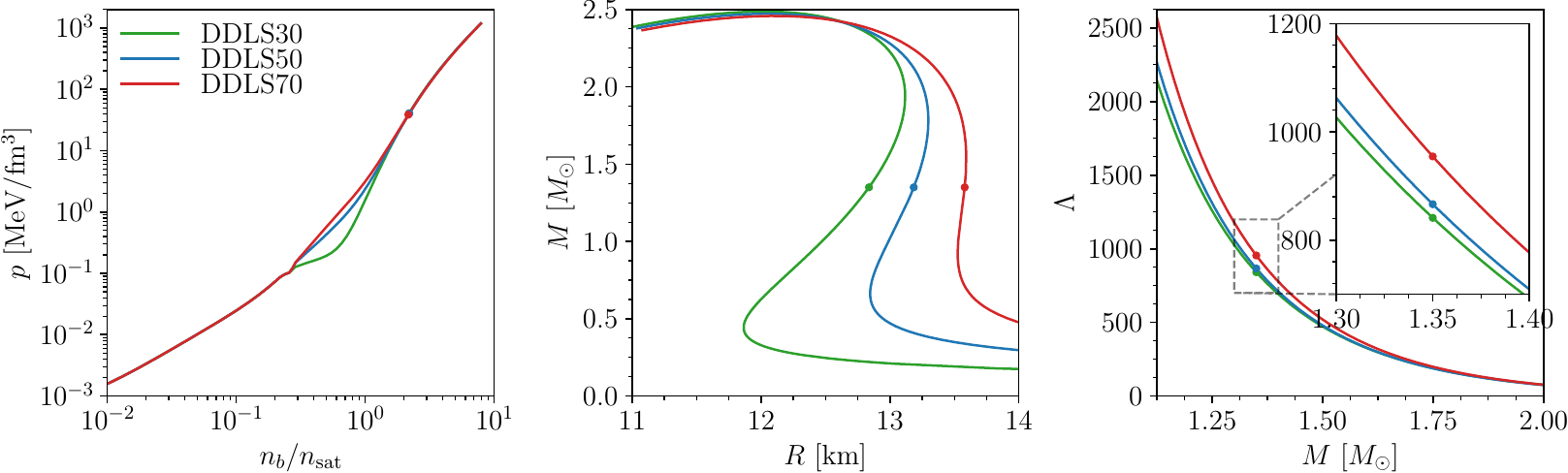}
    \caption{Equations-of-state for cold, $\beta$-equilibrated models (left panel), employed for the construction of the initial data. Mass-radius (middle panel) and tidal deformability-mass (right panel) diagrams for the simulated EOSs. Markers correspond to our simulated configurations.}
    \label{fig:MRL-curves.pdf}
\end{figure*}

In this work, we focus on compact stars composed solely of nucleons. A computationally efficient approach to model the thermodynamics of such stars is provided by the CDF theory of nucleonic matter; see~\cite{Oertel:2016bki,Sedrakian:2023} for reviews. Historically, the application of these models to astrophysical phenomena, particularly to BNS merger simulations, has relied on various heterogeneous parameterizations of the EOS. These parameterizations were often lacking uniform sets that would enable a systematic exploration of the physical properties of nuclear matter. Consequently, there have been no comprehensive studies that examined how variations in the microphysical parameters of the EOS influence the results of merger simulations. To address this issue, Ref.~\cite{Li2023ApJ} introduced a parameterization of the CDF theory based on the density-dependent model DDME2. This new approach allows controlled variations in two key parameters of nuclear matter: the slope of the symmetry energy, $L_{\rm sym}$, and the skewness parameter, $Q_{\rm sat}$ (for definitions, see Eqs. (6) and (7) in Ref.~\cite{Li2023ApJ}). The parameter $L_{\rm sym}$ quantifies the leading-order uncertainty in the symmetry energy, which is strongly correlated with the radius of neutron stars. In contrast, the parameter $Q_{\rm sat}$ characterizes the stiffness (or softness) of symmetric nuclear matter at high densities and is linked to the maximum mass achievable by the compact star sequence for a given parameterization. Attempts have been made to extract the value of $L_{\rm sym}$ from parity-violating electron scattering experiments on $^{208} \mathrm{Pb}$ and $^{40}\mathrm{Ca}$. 
The PREX-II experiment, which analyzed the neutron skin thickness of $^{208}$Pb, suggests a  value of $L_{\text {sym }}=106 \pm 37$ MeV~\cite{PREX:2021,Reed:2021} which is significantly higher than the 
commonly accepted value of $L_{\text {sym }}\simeq 60$~MeV~\cite{Oertel:2016bki}. 
In contrast, the CREX experiment, based on the measurement of neutron skin of $^{48}${Ca}, yields a much lower estimate of $L_{\text {sym }}=20 \pm 30$~MeV~\cite{CREX:2022}. A combined analysis of the PREX and CREX data favors an intermediate value of $L_{\text {sym }} = 52.9 \pm 13.2$ MeV ~\cite{Lattimer:2023} from the overlap region of the two analysis. Additional constraints arise from the analysis of charged pion spectra at high transverse momenta, which imply a range of $42<$ $L_{\text {sym }}<117$~MeV~\cite{SpiRIT:2021}. More recently, heavy-ion collisions at ultrarelativistic energies at the LHC have provided a determination of the neutron skin of $^{208}$Pb, leading to an inferred value of $L_{\text {sym }}=79 \pm 39 \mathrm{MeV}$~\cite{Giacalone:2023}.

Observations of the maximum mass of pulsars have established lower bounds on the value of $Q_{\rm sat}$. However, these limits are sensitive to the assumed composition of the star's matter and vary depending on whether the star is purely nucleonic, contains hyperons, or has a quark matter core. For example, for nucleonic matter, the lower value that is consistent with a two-solar mass non-rotating configuration is $Q_{\rm sat} = -600$~MeV, whereas for hypernuclear stars, this value is $Q_{\rm sat} = 400$~MeV. In the following, the latter value is adopted. 

Finite-temperature extensions of the CDF theory, based on the parameterization introduced in Ref.~\cite{Li2023ApJ}, were developed in Ref.~\cite{Tsiopelas:2024ksy}. These models construct a high-density, finite-temperature equation of state (EOS) that is valid above half the saturation density, seamlessly matched to the low-density EOS from Ref.~\cite{Hempel:2009mc}. The resulting EOS data sets are publicly available in the CompOSE database, https://compose.obspm.fr/eos/xxx, where xxx=322, 324, and  325 for the three nucleonic EOS (used in this work) and xxx = 320, 321, and 323 for the three hyperonic EOS.

Thus, in this work, we employ the nucleonic EOS models DDLS30, DDLS50, and DDLS70 from Ref.~\cite{Tsiopelas:2024ksy}, which correspond to symmetry energy slope values of $L_{\rm sym}=30,50$, and 70 MeV, respectively, at a fixed skewness parameter of $Q_{\rm sat}=400$~MeV. The nuclear saturation density for these models is $n_{\rm sat }=0.152$ fm$^{-3}$.
The cold, $\beta$-equilibrated EOSs used to generate the initial data and determine the properties of isolated neutron stars (NSs) are presented in Fig.~\ref{fig:MRL-curves.pdf}, where the simulated configurations 
($M_1=M_2=1.35 M_{\odot}$) are marked by dots. Here, we assume that the rotation of the initial cold stars can be neglected and these can be approximated as spherically symmetrical, so that it is sufficient to solve the Tolman-Oppenheimer-Volkoff equations. All three EOSs coincide in the high- and low-density limits (left panel), as they share the same low-density EOS from Ref.~\cite{Hempel:2009mc} and high-density value of $Q_{\rm sat}$. The differences emerge in the intermediate density range of $n_b / n_{\rm sat}
=0.25 - 2$ where $n_b$ is the baryonic density. In this regime, DDLS30 is significantly softer, reflecting the lower pressure at saturation density dictated by its smaller $L_{\text {sym }}$ value. Consequently, DDLS30 produces more compact neutron stars (middle panel) and exhibits smaller tidal deformabilities (right panel) compared to DDLS50 and DDLS70. Thus, the selected initial data configuration enables a systematic investigation of the outcomes of BNS simulations as a function of the symmetry energy slope parameter, $L_{\rm sym}$. This approach provides a direct connection between the dynamics and observables of BNS mergers and a fundamental property of nuclear matter. By varying $L_{\rm sym}$, we can explore its impact on crucial aspects such as the stiffness of the EOS, neutron star radii, tidal deformabilities, and the GW signatures emitted during the merger. Consequently, these simulations offer a pathway to constrain nuclear physics models through astrophysical observations, linking microphysical properties of dense matter to macroscopic astrophysical phenomena.

\subsection{Thermal Aspects}\label{sec:thermal}

\begin{figure}[t]
    \centering
    \includegraphics[width=0.9\linewidth]{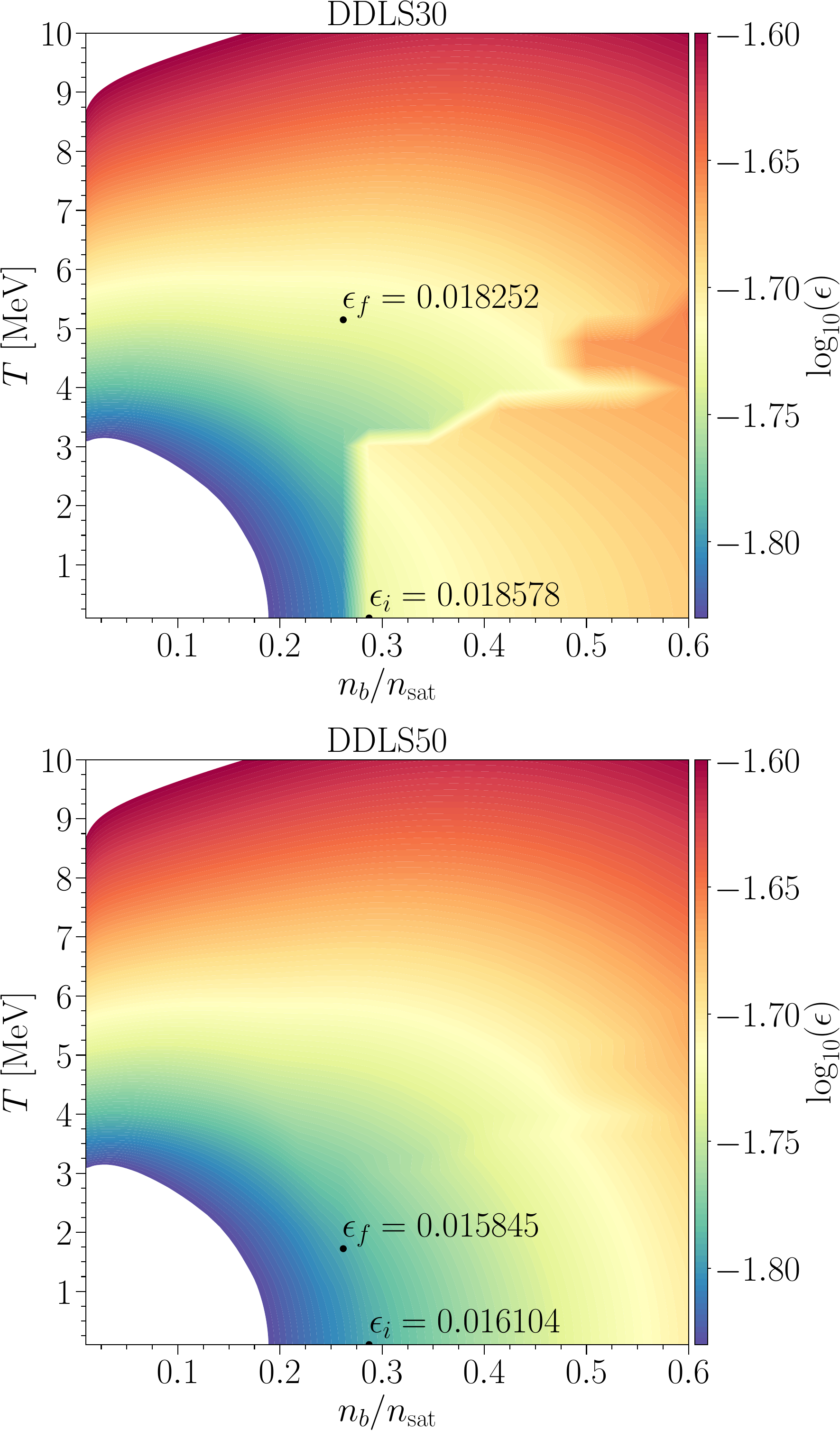}
    \caption{Specific internal energy per baryon $\epsilon$ as a function of scaled density $n_b/n_{\rm sat}$ and temperature $T$ for the ${\rm DDLS}30$ (upper panel), where a clear discontinuity can be seen, and ${\rm DDLS}50$ (lower panel) EOSs at a typical $Y_e = 0.04$, found in outer layers of a NS. The markers labeled as $\epsilon_i$ represent an initial point at temperature $T_i = 0.1~{\rm MeV}$ (e.g., found in the beginning of a simulation) and $n_i/n_{\rm sat}=0.287$, while markers labeled as $\epsilon_f$ are defined with same internal energy as the initial one, but at the immediate predecessor tabulated density $n_f/n_{\rm sat}=0.262$.}
    \label{fig:eps-nb-T}
\end{figure}

Important for the evolution of BNS systems are thermal features encoded in the EOSs, given that structural properties of NSs that imprint astrophysical observables are influenced by temperature. To exemplify some interesting microphysics of the employed EOSs, we show in Fig.~\ref{fig:eps-nb-T} a diagram of the specific internal energy per baryon
\begin{equation}
    \epsilon = \frac{e}{m_bn_b} - 1,
\end{equation}
over a range of densities and temperatures at typical $Y_e = 0.04$ found on the edge of a NS. Here $e$ is the energy density and $m_b$ is a constant mass parameter defined such that $\epsilon \geq 0$ over the whole EOS parameter space.

We note that for the ${\rm DDLS}30$ EOS (upper panel), there is a steep increase in $\epsilon$ for $n_b/n_{\rm sat} \geq 0.25$, coinciding with the region in which the cold, catalyzed counterpart exhibits a softening (see left panel of Fig.~\ref{fig:MRL-curves.pdf}). Such a behavior resembles a first-order phase transition, emerging from the matching of the soft high-density EOS to the low-density EOS of Ref.~\cite{Hempel:2009mc}.

For a qualitative understanding of the effects caused by this energy gradient, consider a point consistent with initial BNS conditions, i.e., with temperature $T_i = 0.1~{\rm MeV}$, electron fraction $Y_e = 0.04$ and a density above the discontinuity $n_i/n_{\rm sat} = 0.287$, hence located close to the NS surface. This point has an internal energy per baryon $\epsilon_i$, as marked in the upper panel of Fig.~\ref{fig:eps-nb-T}. At the beginning of a dynamical evolution, numerical effects commonly drive low-density layers to expand outwards, i.e., crossing the gradient from the right to a point with lower density.
To exemplify the consequence of this energy gradient, we extract from simulation data the specific internal energy (marked as $\epsilon_f$) for $n_f/n_{\rm sat} \approx 0.262$, while $Y_e$ is approximately the same as in the initial point. Then, using the EOS, we invert $T_f = T(\epsilon_f,n_f,Y_e)$ to obtain $T_f \approx 5.15~{\rm MeV}$. At this temperature, the pressure at point $f$ increases to $p_f \approx 1.7~p_i$, thus effectively stiffening the material that expands across the gradient.

As we will see in the following, this behavior has important consequences for our numerical-relativity simulations. In particular, we note the presence of a hot, thermally supported material layer during the inspiral, which modifies at runtime macroscopic properties of the merging NSs, e.g., radius and tidal deformability, and leads to imprints in the observables.

Now, repeating the same procedure for the ${\rm DDLS}50$ EOS (where this property is absent) between the same $i$ and $f$ points as in the ${\rm DDLS}30$ example, the temperature at $f$ is $T_f \approx 1.7~{\rm MeV}$ and the corresponding pressure decreases to $p_f \approx 0.8~p_i$. Therefore, important thermal effects in numerical simulations may arise due to steep gradients in the thermodynamical quantities, as for DDLS$30$, but the effect is milder for DDLS$50$ and DDLS$70$. Importantly, the occurrence of such a gradient in relatively low densities influences the late inspiral and merger, given that the EOS is stiffened.

\begin{table}[h!]
    \centering
        \caption{Properties of the simulated setups. From left to right the columns read simulation name, gravitational mass $M$ in isolation, baryonic mass $M_b$ in isolation, central baryon density $n_c$, coordinate radius $R$, cold tidal deformability $\Lambda$, simulated tidal deformability $\Lambda_{\rm sim}$, number of points per direction in the finest level $n_{\rm fine}$, grid spacing in the finest level $h_{\rm fine}$ and merger time $t_{\rm mrg}$.}
    \begin{tabular}{ccccccccc}
    \toprule
    Simulation  &  $M$ & $M_b$ &$n_c$ & $R$ & $\Lambda$ & $\Lambda_{\rm sim}$ & $n_{\rm fine}$ & $h_{\rm fine}$ \\
     & $[M_\odot]$ & $[M_\odot]$ & $[n_{\rm sat}]$ & $[{\rm km}]$ & & & & $[{\rm m}]$\\
      \hline
     DDLS$30$    &  $1.35$ & $1.48$ & $2.19$ & $12.84$ & $841.2$ & $980.3$& $160$ & $152$ \\
     DDLS$50$    & $1.35$ & $1.47$ & $2.20$ & $13.19$ & $866.3$ &  $876.4$ &$160$ & $156$ \\
     DDLS$70$    & $1.35$ & $1.47$ & $2.16$ & $13.58$ & $954.6$ &  $966.5$ &$160$ & $159$ \\
    \bottomrule
    \end{tabular}
    \label{tab:sim-setups}
\end{table}

\section{Setups and Numerical Methods}
\label{sec:Methods}

\subsection{Initial Data and Simulation Setup}
The BNS setups presented in this work comprise equal-mass, non-rotating NSs with gravitational masses of $M_1 = M_2 = 1.35~M_\odot$. The initial data were constructed using the \sgrid~code~\cite{Tichy:2009yr, Tichy:2012rp, Tichy:2019ouu}. To obtain high-quality data and to avoid biases due to residual eccentricities at merger; cf.~\cite{Foucart:2024kci}, we perform eccentricity reduction following Refs.~\cite{Moldenhauer:2014yaa,Dietrich:2015pxa} and a target eccentricity of $\lesssim 5\times 10^{-3}$. Further details of our simulation setups are summarized in Table~\ref{tab:sim-setups}.

\subsection{Artificial heating through atmosphere treatment}

The standard approach to deal with low-density material inside numerical-relativity simulations is the introduction of an artificial atmosphere to model vacuum and near-vacuum conditions outside the stars, see e.g.~\cite{Rezzolla:2013dea, Yamamoto:2008js, Thierfelder:2011yi, Baiotti:2016qnr}. According to this prescription, material with sufficiently low density (chosen at runtime) is set to floor density and temperature, zero velocity, and the composition is set based on the $\beta$-equilibrium for the considered EOS. This is necessary to ensure numerical stability, as the widely adopted high-resolution shock-capturing (HRSC) scheme and the Valencia Formulation of general relativistic hydrodynamics can face issues for strictly zero densities. More sophisticated alternatives exist, e.g., that of Ref.~\cite{Radice:2013xpa}, which is based on increased dissipation at low densities, but possibly creating artifacts in the ejecta in very dynamical spacetimes.

Connected to the treatment of low-density material, Ref.~\cite{Gittins:2024jui} pointed out that the artificial heating of matter due to shocks with the atmosphere changes the NS structure and increases the tidal deformability of the NSs in a merger simulation. Here, it is worth clarifying that, in our case, besides the known occurrence of such artificial heating in the NS-atmosphere interface, we also have a steep energy gradient in the EOS acting as a source of heating. Therefore, our situation is closer to that reported in Refs.\cite{Gieg:2019yzq, Ujevic:2022nkr}, where heating at relatively high densities was related to first-order phase transitions. In this scenario, the dynamical impacts are lessened due to the dominance of the cold contributions at high densities.

To quantify the thermal effects, we replicate the approach of~\cite{Gittins:2024jui} and compute the tidal deformability at $t = 4~{\rm ms}$, when the stars are still far away but the artificial heating of the stars surface already happened. For this purpose, we first extract the temperatures $T^*$ and densities $n_b^*$ in the orbital plane from our simulation and build a thermal profile $T^* = T^*(n_b^*)$. Then, with the full nuclear EOS tables, we determine the corresponding electron fraction $Y_e^*$ by solving the neutrinoless $\beta$-equilibrium condition
\begin{equation}
    \mu_\Delta(n_b^*,T^*, Y_e^*) = \mu_p + \mu_e - \mu_n = 0,
\end{equation}
which results in a composition profile $Y_e^* = Y_e^*(n_b^*)$. In doing so, we are neglecting the effects of neutrino trapping, given that at this stage, there is no substantial number of neutrinos in the stars. As a next step, from the full EOS tables, we obtain the pressure $p^*(n_b^*) = p(n_b^*, T^*, Y_e^*)$ and specific internal energy per baryon $\epsilon^*(n_b^*) = \epsilon(n_b^*, T^*, Y_e^*)$. Finally, we solve the TOV equations using the one-dimensional EOS $(p^*(n_b^*), \epsilon^*(n_b^*))$ and evaluate the simulated tidal deformability $\Lambda_{\rm sim} = \Lambda^*(M_b)$ at the same baryonic mass as the initial condition.

As shown in Table~\ref{tab:sim-setups}, under simulation conditions, the tidal deformability increases for DDLS$30$ by $\approx 17\%$, while it increases modestly for DDLS$50$ and DDLS$70$ on the $\approx 1\%$ level. In fact, DDLS$30$ ends up with an effective tidal deformability that is $2\%$ larger than the stiffest DDLS$70$. 

\begin{figure}[htpb!]
    \centering
    \includegraphics[width=0.9\linewidth]{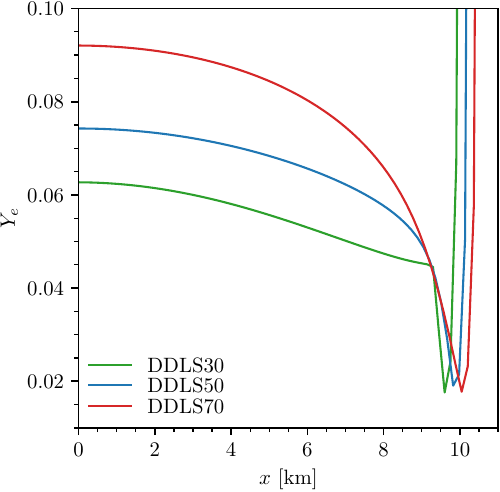}
    \caption{Compositional profile of the NSs in the initial data along the $x$-axis, where $x=0$ coincides with the center of the NSs, while transition to the artificial atmosphere is seen in the increase of $Y_e$.}
    \label{fig:ID-profile.pdf}
\end{figure}

Besides the structural differences between NSs produced with those models, one interesting consequence of increased $L_{\rm sym}$ is that cold, $\beta$-equilibrated configurations are typically more protonized, as smaller isospin asymmetry is energetically favored. This can be seen by considering that, at zero temperature and expanding the energy per baryon up to second order in the isospin asymmetry $\delta = 1-2Y_p$~\cite{Li:2023bid}, the difference between the chemical potentials of neutrons and protons is (see Appendix B of Ref.~\cite{Most:2021ktk} for a derivation)
\begin{eqnarray}
    \hat{\mu} &=& \mu_n - \mu_p = 4(1-2Y_p)E_{\rm sym}(n).
\end{eqnarray}
Since the symmetry energy $E_{\rm sym}(n)$ grows linearly with $L_{\rm sym}$, so does $\hat{\mu}$. Then, neutrinoless $\beta$-equilibrium implies that the electron chemical potential satisfies $\mu_e = \hat\mu$, meaning that higher $L_{\rm sym}$ leads to higher electron fraction $Y_e$, thus higher proton fraction $Y_p$ is required for charge neutrality.
This is what we verify in Fig.~\ref{fig:ID-profile.pdf}, where we show the compositional profile of the NSs at the initial time slice.

\subsection{Dynamical Evolution}\label{sec:dyn-evo}

The evolution of matter and spacetime fields is performed with the \bam~code~\cite{Bruegmann:2006ulg, Thierfelder:2011yi, Bernuzzi:2016pie, Gieg:2022,Schianchi:2023uky}, whose computational domain is composed of a hierarchy of nested Cartesian levels (7 in total for our simulations) with a 2:1 refinement strategy for the grid spacing. Adaptive mesh refinement allows a number of innermost levels to follow the motion of the stars, while refluxing via conservative mesh refinement~\cite{Dietrich:2015iva} is employed to improve mass conservation across refinement boundaries.

Geometry fields are evolved using the Z4c formalism~\cite{Hilditch:2012fp,Bernuzzi:2009ex} with fourth-order spatial discretization for derivatives. The moving punctures method~\cite{Campanelli:2005dd, Bruegmann:2006ulg} is adopted, i.e., gauge fields evolve according to the $1 +\log$ slicing~\cite{Bona:1994dr} and the Gamma-driver conditions~\cite{Alcubierre:2002kk}. 

For the matter fields, we employ the HRSC scheme, in which primitive fluid variables are computed at cell interfaces via WENOZ reconstruction~\cite{Borges:2008} and the HLL Riemann solver~\cite{Harten:1983, Toro:1999} is used to evaluate inter-cell fluxes. As previously stated, very low density regions are treated as a cold, static, $\beta$-equilibrated atmosphere.

Both matter and spacetime quantities are evolved in time by the method-of-lines with an explicit fourth-order Runge-Kutta integrator and Berger-Oliger time stepper~\cite{Berger:1984}.

We also include neutrino transport in our simulations with the M1 scheme~\cite{Shibata:2011kx,Foucart:2015gaa,Foucart:2016rxm,Weih:2020wpo,Radice:2021jtw,Musolino:2023pao}, following the implementation of Ref.~\cite{Schianchi:2023uky}, with the same sets of reactions, except for the addition of elastic scattering of neutrinos on alpha particles. 

\section{Simulation Results}
\label{sec:Results} 

\begin{figure}[htpb!]
    \centering
    \includegraphics[width=0.85\linewidth]{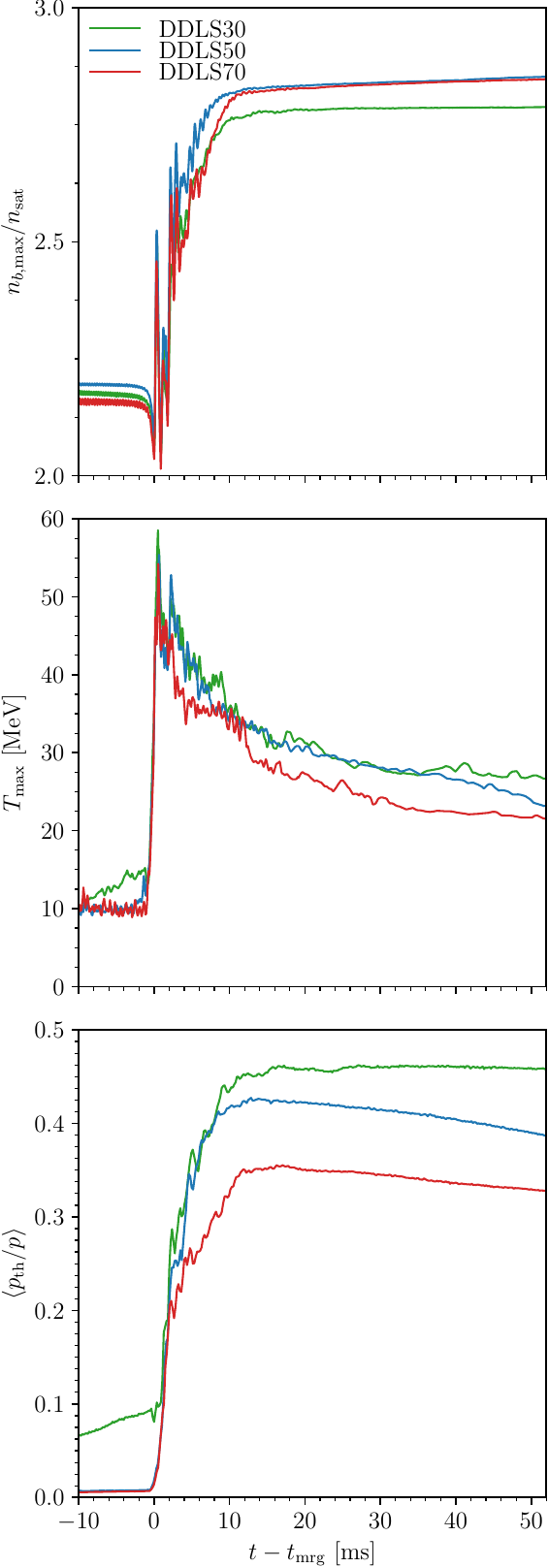}
    \caption{Evolution of the central rest-mass density (upper panel), maximum temperature (middle panel) and ratio of mass-averaged thermal-to-total pressure (lower panel) of matter elements with $n_b/n_{\rm sat} = [0.1,2.0]$, roughly corresponding to the the hot annulus formed in the post-merger. A moving average filter with a $0.2~{\rm ms}$ time window is employed.}
    \label{fig:rhomax}
\end{figure}

\subsection{Merger and Post-Merger Dynamics}
All our simulations start at an initial coordinate separation $d\approx 46.9~{\rm km}$, merging after $\approx 6$ orbits, with no noticeable difference in the inspiral dynamics. For convenience, we present our results shifted by the merger time $t_{\rm mrg}$ and focus on the post-merger stage up to the end of our simulations at $t - t_{\rm mrg} \approx 52~{\rm ms}$.

In Fig.~\ref{fig:rhomax}, we present the evolution of thermodynamical quantities of interest. The upper panel represents the evolution of the central density, where the merger is marked by a rapid increase due to matter compression, followed by oscillations that are dampened due to the emission of GWs on a timescale of $\approx 10~{\rm ms}$. Afterwards, the densest portion of the remnants reaches quasi-equilibrium states with approximately constant density. We note that $n_{b,\rm max}$ of the stiffest DDLS$70$ evolves more slowly into approximately the same value as DDLS$50$, while the remnant of DDLS$30$ has a smaller central density by about $2.3\%$. This effect can be traced back to the evolution of the maximum temperature (middle panel), where the softest DDLS$30$ shows the highest temperature before the merger, due to the thermal features of DDLS$30$ discussed in Sec.~\ref{sec:thermal}, but also during the merger, followed by a steady cooling via neutrino emission over a timescale of a few tens of ms. Not surprisingly, DDLS$70$ exhibits the smallest temperature increase and, accordingly, evolves into a cooler remnant. On the other hand, DDLS$50$ has maximum temperatures close to DDLS$30$ along most of the merger aftermath, reaching intermediate temperatures by the end of our simulation.

\begin{figure*}[t]
    \centering
    \includegraphics[width=\linewidth]{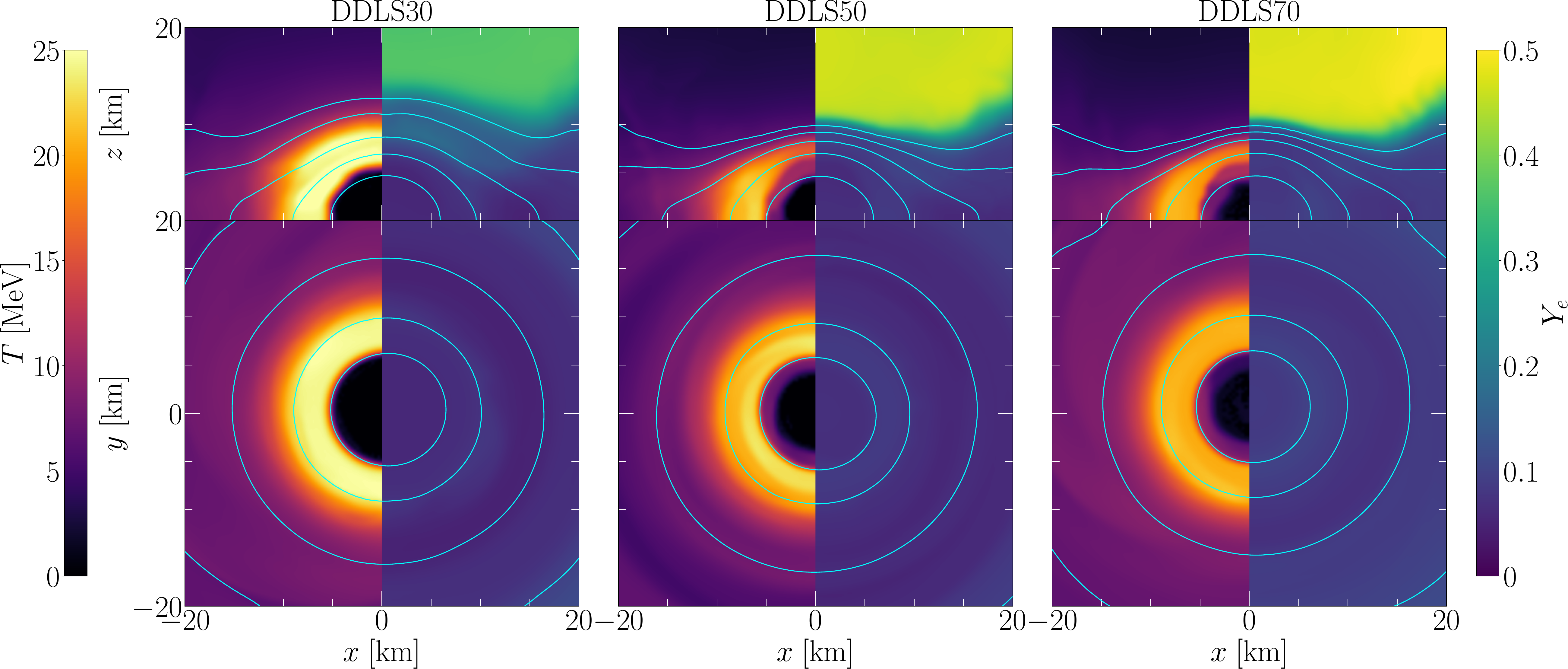}
    \caption{Temperature (left half) and electron fraction (right half) of the remnants for simulations DDLS$30$ (left panel), DDLS$50$ (middle panel) and DDLS$70$ (right panel), on the $x-y$ plane (bottom panels) and $x-z$ plane (top panels) at the end of the simulations $t-t_{\rm mrg}\approx 52~{\rm ms}$. Contour lines mark isodensity surfaces with $n_b/n_{\rm sat} = [10^{-3}, 10^{-2}, 10^{-1}, 10^0, 2]$. The softer DDLS$30$ EOS produces the hottest annular region between one and two saturation densities, followed by DDLS$50$ and DDLS$70$, respectively, with the least electronized remnant (see text). Likewise, the low density, comparatively colder polar region of all remnants exhibits a sustained neutrino-wind with increased $Y_e$ for higher $L_{\rm sym}$.}
    \label{fig:2d-panel}
\end{figure*}

Therefore, one expects that thermal effects are more pronounced for the evolution of the remnants of DDLS$30$ and DDLS$50$. To quantify those effects, similarly to Ref.~\cite{Bauswein:2010dn} and many others, we define the thermal pressure $p_{\rm th}$ of a matter element as
\begin{equation}\label{eq:pth}
    p_{\rm th}(n_b, T, Y_e) = p(n_b,T,Y_e) - p(n_b,T_0, Y_e),
\end{equation}
where $p(n_b,T,Y_e)$ is the pressure obtained from the EOS table with simulated matter state $(n_b, T, Y_e)$, while $T_0 = 0.1~{\rm MeV}$ corresponds to the minimum tabulated temperature, thus $p(n_b,T_0,Y_e)$ represents the ``cold'' contribution to the pressure. Typically, the pressure derived from nuclear EOSs at high densities is rather insensitive to temperature, but thermal support is generally important at smaller densities, e.g., below $n_{\rm sat}$. Hence, we specify the following analysis to the hot annular region (such as, for instance, the ones depicted in Fig.~\ref{fig:2d-panel}), where the bulk of hot material is found in a region that spans densities in the range $n_b/n_{\rm sat} = [0.1,~2]$. Next, we define the mass-averaged ratio of thermal-to-total pressure on the orbital plane as
\begin{equation}
    \langle p_{\rm th}/p\rangle = \frac{\int_{\mathcal{R}}~(p_{\rm th}/p) \sqrt{\gamma}D~dxdy }{\int_{\mathcal{R}} \sqrt{\gamma}D~{dxdy}},
\end{equation}
where $D = W m_b n_b$ is the rest-mass density of a fluid element as seen by Eulerian observers, $W$ is the corresponding Lorentz factor, $\gamma$ is the determinant of the 3-metric and $\mathcal{R}$ is the region of the $x-y$ plane encompassing the hot annulus, where $0.1 \leq n_b/n_{\rm sat} \leq 2.0$. The mass-averaged ratio of thermal-to-total pressure is depicted in the lower panel of Fig.~\ref{fig:rhomax}, where we can see that thermal support plays an important role for DDLS$30$ prior to the merger, while it is comparable for DDLS$30$ and DDLS$50$ in the first $10~{\rm ms}$ of the post-merger. Interestingly, although the evolution of the maximum temperature is similar for those runs, over longer timescales $\langle p_{\rm th}/p \rangle$ remains above the $45\%$ level for DDLS$30$, suggesting that thermal effects are more important for the evolution of the remnant governed by this EOS, in particular by preventing material with $n_b/n_{\rm sat} \leq 2$ from falling into the core and increasing  $n_{b,\rm max}$. Contrary, for DDLS$70$, whose maximum temperature remains lower, thermal support is less prevalent and the remnant eventually reaches equilibrium at a higher density.

In Fig.~\ref{fig:2d-panel}, we present the thermodynamical state of the remnants by the end of our simulations, namely the temperature (left half) and electron fraction (right half) on the $x-y$ (orbital) plane (bottom panels) and $x-z$ plane (top panels), with isodensity contours in cyan.  In all cases, the hot annulus in the orbital plane is restricted between one and two saturation densities, and, as expected, temperatures are ordered according to the softness of the EOS. The more pronounced differences in stiffness across the models between $(0.25 - 1) n_{\rm sat}$ (see right panel of Fig.~\ref{fig:MRL-curves.pdf}) has no significant impact in the compactness of the disks. However, we see that the hot layer is particularly expanded in the $x-z$ plane for the DDLS$30$ run, where a thicker bulge around the polar cap is present, with lower density material supported by thermal pressure.

It is also worth noting that the electron fraction distributions are considerably different within the remnant (hereafter defined as matter with $m_bn_b \geq 10^{13}~{\rm g/cm^3}$, or $n_b \geq 0.039~n_{\rm sat}$), as summarized in Table~\ref{tab:rem-ej-prop}, where we note that the average electron fraction in the remnant $\langle Y_e\rangle_{\rm rem}$ follows the trend anticipated by the initial data, i.e., that it increases with $L_{\rm sym}$, being ${\rm DDLS}30$ the most neutron-rich. In this case, the production of electron neutrinos $\nu_e$ via electron capture on proton $e^- + p \rightarrow n+\nu_e$ is limited by the reduced availability of electrons and protons in the remnant (see Fig.~\ref{fig:ID-profile.pdf}). Consequently, a smaller absorption of $\nu_e$ in the outflows (importantly produced by neutrino irradiation at this late stage) leads to an overall less leptonized ejecta, as seen in the upper panels of Fig.~\ref{fig:2d-panel} and confirmed in the smaller average electron fraction of the ejecta $\langle Y_e\rangle_{\rm ej}$ presented in the Table~\ref{tab:rem-ej-prop}.

\begin{table}[t]
    \centering
        \caption{Average thermodynamical properties of the remnant and ejecta by the end of our simulations. From left to right the columns read simulation name, remnant baryonic mass, ejecta mass, average electron fraction of the remnant, average electron fraction of the ejecta and average asymptotic velocity of the ejecta.}
    \begin{tabular}{cccccc}
    \toprule
    Simulation  &  $M_{\rm rem}$ & $M_{\rm ej}$ & $\langle Y_e\rangle_{\rm rem}$ & $\langle Y_e \rangle_{\rm ej}$ & $\langle v_\infty\rangle$ \\
     & $[M_\odot]$ & $[10^{-2} M_\odot]$ & & & \\
      \hline
     DDLS$30$    &  $2.76$ & $0.28$ & $0.068$ & $0.24$ & $0.16$  \\
     DDLS$50$    & $2.74$ & $0.25$ & $0.075$ & $0.28$ & $0.15$  \\
     DDLS$70$    & $2.76$ & $0.22$ & $0.087$ & $0.31$ & $0.14$ \\
    \bottomrule
    \end{tabular}
    \label{tab:rem-ej-prop}
\end{table}

\begin{figure*}[t]
    \centering
    \includegraphics[width=0.8\linewidth]{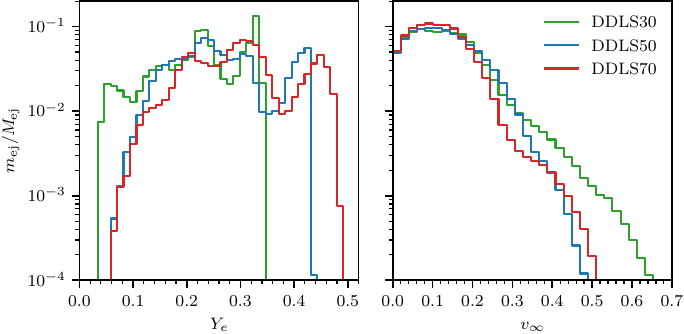}
    \caption{Mass histograms of electron fraction (left panel) and asymptotic velocity (right panel) for the entire simulations.}
    \label{fig:Ye-hist}
\end{figure*}

Finally, in Fig.~\ref{fig:Ye-hist} we show the mass histograms of electron fraction (left panel) and asymptotic velocity (right panel) in the ejecta of our runs. The distributions for ${\rm DDLS}50$ and ${\rm DDLS}70$ share remarkably similar features, in the velocities and electron fraction. In both cases, there is a small fraction of neutron-rich ejecta $Y_e \lesssim 0.1$, with peaks in the interval $Y_e = 0.2-0.4$ and a decay at higher electron fractions. Noteworthy is that ${\rm DDLS}70$ achieves higher $Y_e$ values at the tail of the distribution, in agreement with the aforementioned leptonization process. On the other hand, the ${\rm DDLS}30$ run exhibits a higher amount of neutron-rich material, which we believe to be consistent with the sizable amount of high velocity ejecta $v_\infty \geq 0.4$ produced by the collision of hot, thermally supported material around the merger and early post-merger. Hence, fast escaping matter do not efficiently leptonize before reaching the detection sphere, where ejecta properties are extracted.

Here we make a remark about the mass and velocity of the ejecta found in DDLS$30$, where we remind that the simulation tidal deformability is higher, due to the thermal stiffening of the EOS at intermediate densities, with a corresponding hot layer visible throughout the inspiral (see the higher temperature depicted in Fig.~\ref{fig:rhomax}). Such a structure resembles a standing shock, located at the positions where the steep energy gradient may be found. Therefore, additionally to the shocks taking place at the merger and early post-merger (mainly produced by shear layer interaction and core bounce)~\cite{Radice:2018pdn, Rosswog:2024vfe}, which are responsible for releasing most of the dynamical outflows, we postulate that energy gradients in the EOS may act in a similar manner in driving the outflows. It seems, however, non-trivial to disentangle the contributions from this phenomena and the hydrodynamical shocks following the merger.

\subsection{Gravitational-Wave Signal}

In this section, we analyze the GW signatures produced by the different merging systems. The waveforms are extracted at a finite radius $r = 1000~M_\odot \approx 1477~{\rm km}$.

The GW signal can be split into three consecutive stages: (i) the inspiral: as the evolution progresses, the binary loses energy in the form of GWs, the orbital radius diminishes, and the GW frequency increases, leading to the characteristic chirp signal. Once the stars are sufficiently close, tidal interactions, described by a combined tidal deformability $\tilde{\Lambda}$~\cite{Hinderer:2009ca} play a significant role and introduce dephasings among the models. (ii) Merger stage: defined by the maximal amplitude of the $(2,2)$ mode, and (iii) post-merger stage, where GW emission is set by the motion of the remnant. Typically, during the early post-merger, the GW amplitude decreases on a timescale of tens of ms, due to the evolution of the remnant towards an axisymmetric configuration. Again, energy losses via gravitational radiation damp oscillatory motions within the remnant, which, in turn, decrease the GW amplitude.

With this picture in mind, the prospect of connecting GW observables to microscopic properties of dense matter encoded in the EOS is promising.

\begin{figure}[t]
    \centering
    \includegraphics[width=\linewidth]{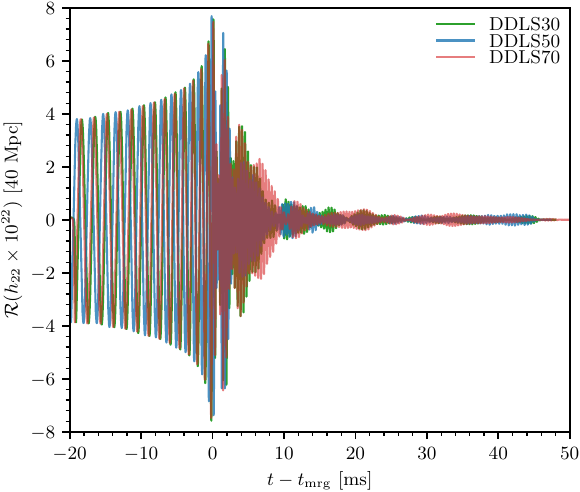}
    \caption{Evolution of the real part for the (2,2)-mode of the GW $h$, at a distance $40$ Mpc. The merger times are $t_{\rm mrg} = \{19.21, 19.91, 19.37\} ~{\rm ms}$ for DDSL30, DDSL50 and DDSL70, respectively. The increasing merger times follow the ordering of simulated tidal deformabilities.}
    \label{fig:waveform22}
\end{figure}

To this end, in Fig.~\ref{fig:waveform22} we show the time evolution of the real part of the strain (for the dominant $(2,2)$ mode), where the vertical lines indicate the merger times $t_{\rm mrg}$ for each run. We notice that differences in the waveforms arise close to the merger and are more evident along the post-merger stage. Also, the merging times are ordered with increased (simulated) tidal deformability, since more deformable (less compact) NSs come into contact earlier. It is interesting to note that due to the dominance of thermal effects in the DDLS$30$ run, its simulated tidal deformability becomes similar to that of DDLS$70$, hence the signals evolve rather similarly. Contrary, the more compact DDLS$50$ setup merges $\approx 0.6~{\rm ms}$ later than DDLS$70$. 

\begin{figure*}[htpb!]
    \centering
    \includegraphics[width=\linewidth]{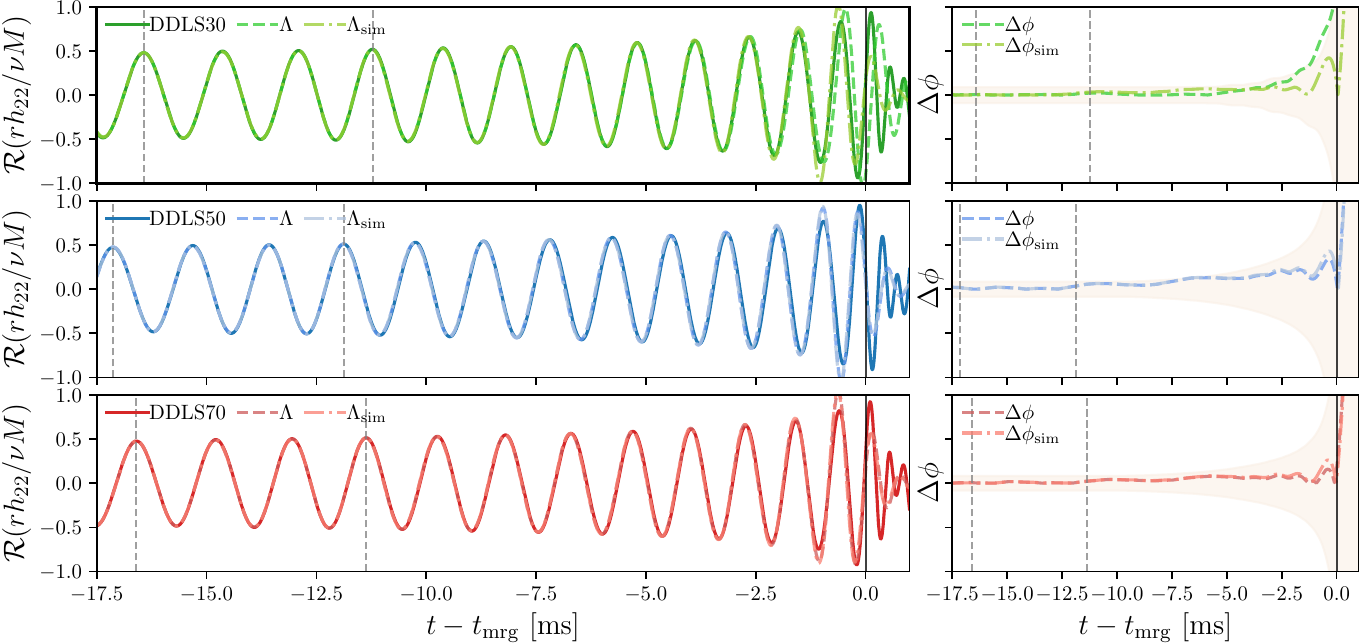}
    \caption{Comparison between the waveforms (left panels) extracted from the simulations (thick lines) and the waveforms obtained with the {\tt IMPhenomXAS\_NRTidalv3} model, either using the static $\Lambda$ (dashed lines) or the estimated $\Lambda_{\rm sim}$ (dash-dotted lines). The agreement between numerical and approximant waveforms is quantified by the dephasings (right panels), with uncertainty bands depicted as shaded areas.}
    \label{fig:ntidalv3}
\end{figure*}

Now, in order to assess whether the differences in the ispiral-merger waveforms can be decisively traced to differences in $L_{\rm sym}$, we present the scaled strain of the dominant $(2,2)$ mode for the inspiral stage of our simulations in the left panels of Fig.~\ref{fig:ntidalv3}, where thick lines represent numerical waveforms extracted from our simulations, compared to waveforms computed using the {\tt IMPhenomXAS\_NRTidalv3} approximant~\cite{Abac:2023ujg} either with the static $\Lambda$ (dashed lines) or the simulated $\Lambda_{\rm sim}$ (dash-dotted lines). All of the waveforms are aligned in time within the intervals indicated by vertical dashed lines, and are depicted up to a few ${\rm ms}$ after the merger. There, we note that the waveform approximant satisfactorily reproduces the numerical waveforms, in particular during the inspiral. In order to quantify the differences between numerical and approximant waveforms, we introduce the dephasings
\begin{eqnarray}
    \Delta \phi &=& \phi_{\rm NR} - \phi_{\rm NRTidalv3}(\Lambda), \\
    \Delta \phi_{\rm sim} &=& \phi_{\rm NR} - \phi_{\rm NRTidalv3}(\Lambda_{\rm sim}),
\end{eqnarray}
depicted in the right panels of Fig.~\ref{fig:ntidalv3} as dashed and dash-dotted lines, respectively, along with the error bands (shaded regions). The error band, $\epsilon$, is estimated using two terms, i.e. $\epsilon^2 = {\rm max}[\epsilon^2_{\rm Res} + \epsilon^2_{\rm Rad}]$, where ``max" indicates that we are keeping the highest value as time passes by to ensure a monotonically increasing uncertainty. In this relation, $\epsilon_{\rm Res}$ is obtained through the phase difference between the highest and medium resolution at a given extraction radius, and $\epsilon_{\rm Rad}$ is obtained by calculating the phase difference between two different extraction radii in the highest resolution available. In our case, we consider the radii 600~$M_\odot$ and 1000~$M_\odot$. 

Overall, the early inspiral $t - t_{\rm mrg} \leq -2.5~{\rm ms}$ signal exhibits dephasings well within the uncertainty bands. Closer to the merger $t - t_{\rm mrg} \geq -2.5~{\rm ms}$, the dephasings for DDLS50 and DDLS70 are comparable, either with the static $\Lambda$ or the estimated $\Lambda_{\rm sim}$. Noteworthy is the case of DDLS30, in which the approximant with static tidal deformability grows outside the error band, but the adoption of the simulated tidal deformability $\Lambda_{\rm sim}$, which captures the dominance of thermal effects, improves the agreement between NR and approximant waveforms, as the dephasing is driven to within the uncertainty band.

In addition, the fact that the {\tt IMPhenomXAS\_NRTidalv3} approximant nicely reproduces our NR waveforms reinforces that most of the differences in the inspiral waveform induced by different $L_{\rm sym}$ are encoded in the macroscopic parameters of the NSs, namely tidal deformability and compactness. Conversely, different EOSs producing NSs with similar compactness and tidal deformability as the ones simulated in this work would, then, lead to a quantitatively similar GW signal.

\begin{figure}[t]
    \centering
    \includegraphics[width=\linewidth]{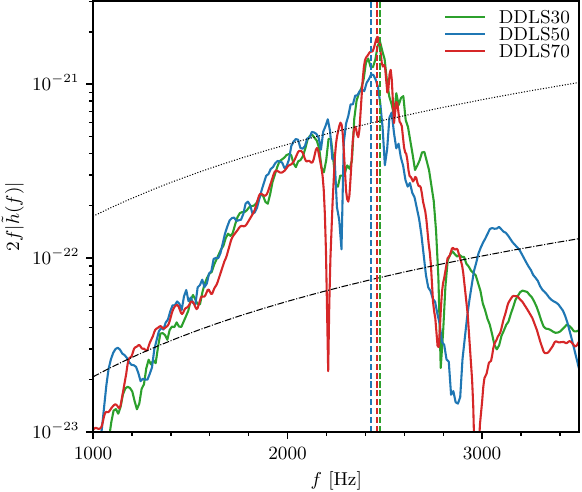}
    \caption{GW spectra at $40$ Mpc for the post-merger signal, extracted $1.5~{\rm ms}$ after the merger and at a finite coordinate radius $r = 1000~M_\odot$. The dominant frequency $f_2 = \{2.475, 2.429, 2.461\}~{\rm kHz}$ for DDSL30, DDSL50 and DDSL70, respectively, are marked with vertical lines. Black dotted (dash-dotted) lines correspond to the design sensitivity of advanced LIGO~\cite{LIGOScientific:2014pky} (Einstein Telescope~\cite{Hild:2010id}).}
    \label{fig:GW-spectrum.pdf}
\end{figure}

Regarding the post-merger signal, in Fig.~\ref{fig:GW-spectrum.pdf} we show the GW spectra computed $1.5~{\rm ms}$ after the merger, including the $l=2,3$ modes . Vertical lines mark the post-merger peaks $f_2$, commonly associated to quadrupolar oscillations of the remnant~\cite{Stergioulas:2011gd}. For the highest resolution runs, the peak frequencies are shown in Table~\ref{tab:peak-freq} for DDLS30, DDLS50, and DDLS70. On the other hand, the peak frequencies obtained with the fitting formula Eq.~(33) of Ref.~\cite{Vretinaris:2019spn}, for which $f_{2, {\rm fit}}$ is given in terms of the chirp mass $M_{\rm chirp}$ and the effective tidal deformability $\tilde{\Lambda}$ (which is equal to the individual $\Lambda$ for our symmetric systems) are also shown in Table~\ref{tab:peak-freq}. There the fitted frequency $f_{2,{\rm fit}}(\Lambda)$ ($f_{2,{\rm fit}}(\Lambda_{\rm sim})$) is computed with the isolated (simulated) tidal deformability and the peak frequency shifts are $\Delta f_2(\Lambda) = f_2 - f_{2,{\rm fit}}(\Lambda)$, and $\Delta f_2(\Lambda_{\rm sim}) = f_2 - f_{2,{\rm fit}}(\Lambda_{\rm sim})$.

Overall, we note that the frequency shifts are smaller than the maximum reported residual for the fitting formula $\sim 0.302~{\rm kHz}$. Hence, the post-merger peak frequencies are well captured by the fit, which depends only on macroscopic binary parameters. Moreover, we also note that the phase shifts are smaller when the simulated tidal deformability is employed to compute the peak frequency, again reinforcing that thermal effects arising in numerical simulations have a sizable effect on GW properties, as anticipated by Ref.~\cite{Gittins:2024jui}. This is more evident in the DDLS30 case. 

It is worth pointing out that the difference in peak frequency $f_2$ for different $L_{\rm sym}$ is very small (at most $\sim 50~{\rm Hz}$), hindering the distinction of our EOSs based on $f_2$. Interestingly, our findings are in good qualitative agreement with Ref.~\cite{Kochankovski:2025lqc}, where they also report similar peak frequencies for all DDLS nucleonic runs. Therefore, our results indicate that the main role of varying $L_{\rm sym}$ in the post-merger GW spectrum is encoded in variations of $\Lambda$, which, on its turn, is degenerate with respect to other nuclear parameters or even EOS models. Moreover, we performed two additional simulations with lower resolutions for each $L$, in particular we use (254~m, 381~m), (260~m, 390~m) and (266~m, 399~m) for DDLS30, DDLS50 and DDLS70, respectively. We found that the largest difference in the main frequency peak between two different simulations (irrespective to resolution and $L_{\rm sym}$) is less than $95$~Hz.

\begin{table}[t]
    \centering
        \caption{Post-merger peak frequencies. From left to right the columns read simulation name, fitted value for the static $\Lambda$, fitted value for the simulated $\Lambda_{\rm sim}$, extracted value from our simulations, frequency shift with respect to the fit using $\Lambda$, and frequency shift with respect to the fit using $\Lambda_{\rm sim}$.}
    \begin{tabular}{cccccc}
    \toprule
    Simulation  &  $f_{2,{\rm fit}}(\Lambda)$ & $f_{2,{\rm fit}}(\Lambda_{\rm sim})$ &$f_2$ & $\Delta f_2 (\Lambda)$ & $\Delta f_2(\Lambda_{\rm sim})$\\
     & $[{\rm kHz}]$ & $[{\rm kHz}]$ & $[{\rm kHz}]$& $[{\rm kHz}]$ & $[{\rm kHz}]$\\
      \hline
     DDLS$30$    &  $2.593$ & $2.481$ & $2.475$ & $-0.118$ & $-0.006$\\
     DDLS$50$    & $2.571$ & $2.562$ & $2.429$ & $-0.142$ & $-0.133$\\
     DDLS$70$    & $2.500$ & $2.491$ & $2.461$ & $-0.039$ & $-0.030$\\
    \bottomrule
    \end{tabular}
    \label{tab:peak-freq}
\end{table}

\subsection{Abundances}
\begin{figure}[t]
    \centering
\includegraphics[width=\linewidth]{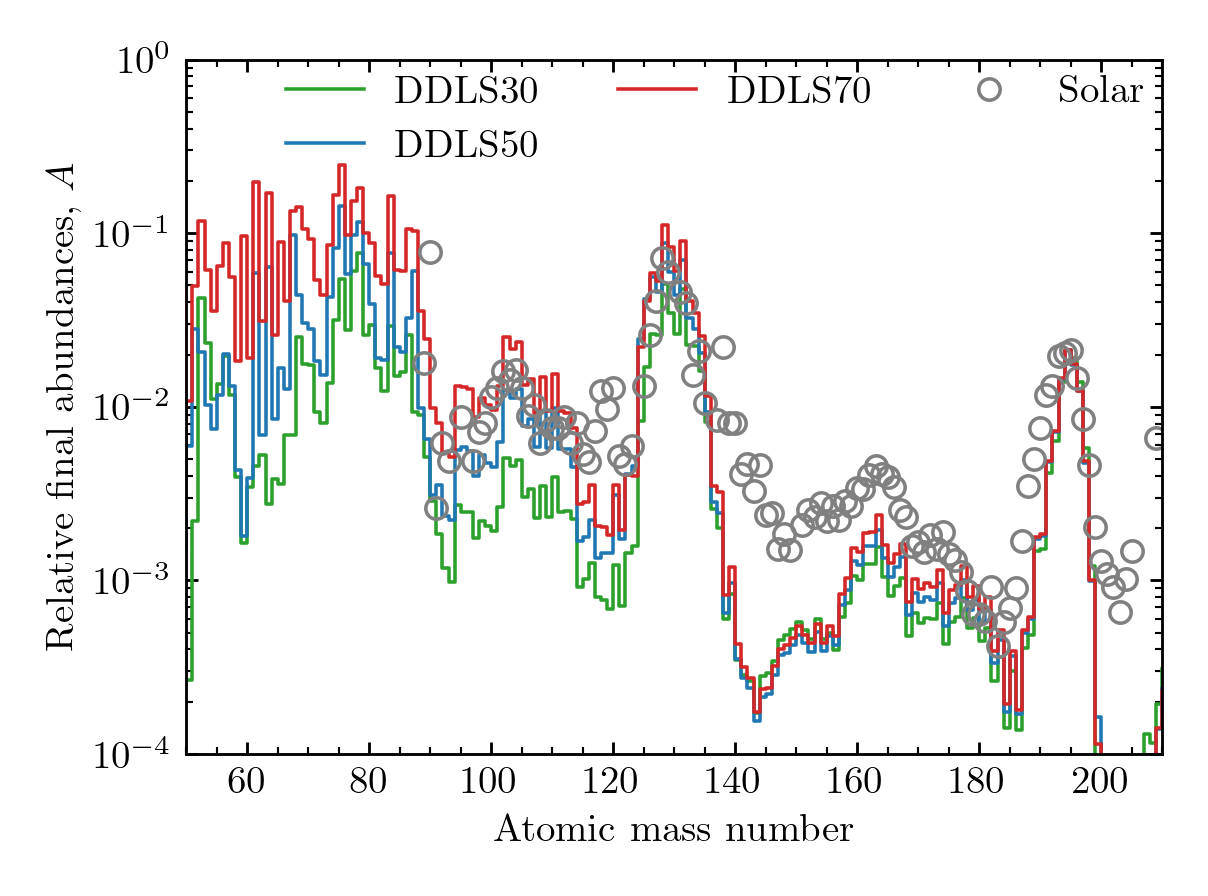}
    \caption{Nucleosynthetic yields of our simulations, with relative abundances normalized to the solar value at $A = 195$.}
    \label{fig:rp-yields}
\end{figure}

Motivated by the quantitative differences observed in the ejecta properties (Sec.~\ref{sec:dyn-evo}), we now investigate the abundances of r-process elements estimated with our simulation data. To this end, we follow the procedure of Refs.~\cite{Radice:2018pdn, Radice:2016dwd,Schianchi:2023uky} and employ the \texttt{Skynet} nuclear reaction network~\cite{Lippuner:2015gwa} to compute the nucleosynthetic yields produced in the ejecta. The relative abundances of the different isotopes are shown with respect to the mass number $A$ in Fig.~\ref{fig:rp-yields}. As standard, the relative abundances are normalized such that, across different simulations, at $A = 195$ the abundances match the solar one~\cite{Prantzos2020}. 

The abundances are broadly consistent across the different EoSs for heavier elements $A \gtrsim 120$, nicely reproducing the solar abundances in the second r-process peak, but with underproduction in the rare-earth peak (around $A = 160$). Most notable is the increasing fractions of lighter elements $A \lesssim 100$ for DDLS30, DDLS50, and DDLS70, respectively, which follow from the increasing electron fraction in the respective ejecta. 

This suggests that the kilonova associated with our simulations could, in principle, distinguish the EoSs, as details of the elemental distribution govern the opacity of the medium for photons~\cite{Lippuner:2015gwa, Roberts:2011, Barnes:2013wka, Tanaka:2013ana}. On the other hand, the properties of the ejecta (mass, electron fraction, specific entropy, and velocity), whose averages differ in our simulations by as much as $\sim 30\%$, importantly impact the kilonova signal~\cite{Bulla:2019muo, Kasen:2017sxr, Perego:2017wtu}. Hence, future work targeting a characterization of the electromagnetic counterparts will be needed in order to clarify if there are sufficient imprints to distinguish $L_{\rm sym}$, and if present, whether it is possible to convincingly correlate such signatures to $L_{\rm sym}$.

Finally, we remark that it was not possible to reliably estimate kilonova lightcurves for the present simulations based on the methods of Ref.~\cite{Schianchi:2023uky}, because our grid setup was not sufficiently resolved to capture enough unbound material at the homologous expansion phase (typically around $t - t_{\rm mrg} \sim 30~{\rm ms}$~\cite{Neuweiler:2022eum}). On the other hand, sufficient unbound mass was only found at an earlier stage $t - t_{\rm mrg} \sim 6~{\rm ms}$, where dynamical outflows are too neutron-rich and opacities are uncertain~\cite{Bulla:2022mwo}. Given these restrictions, we decided to not provide estimated kilonova lightcurves which potentially could have led to wrong conclusions.

\section{Discussion and Conclusions}
\label{sec:Conclusions}

In this work, we have performed a set of BNS merger simulations aiming to clarify whether slope of the symmetry energy may leave imprints in GWs or along the post-merger evolution. To do so, we employed a set of finite-temperature, composition-dependent EOSs, constructed to systematically explore the variations of $L_{\rm sym} = \{30, 50, 70\}~{\rm MeV}$ in a controlled manner.

Our main findings may be divided into two parts. The first is focused on the artificial heating experienced by matter along a numerical simulation. It is currently well-understood that shocks in the interface between NS and atmosphere are responsible for artificially heating low-density material (see Ref.~\cite{Hammond:2021vtv} for a review, Ref.~\cite{Gittins:2024jui} for an analysis of the observable consequences, and Ref.~\cite{Doulis:2022vkx, Doulis:2024aew}, where the adoption of an entropy-based flux scheme conclusively points to the phenomena). Added to that, we observed that heating may also be produced by steep gradients in the EOS (Fig.~\ref{fig:eps-nb-T}), which is in agreement with previous studies where first-order phase transitions were considered. Contrary, however, to those studies, where increased thermal contributions were restricted to high densities, our DDLS$30$ simulation showcased
a similar behavior at lower densities. We observe that  higher temperatures play a significant role, as thermal support in the range $n_b = (0.1 - 1)~n_{\rm sat}$ effectively stiffens the EOS, impacting the resulting tidal deformability and compactness of the NSs. Furthermore, for setups with effective stiffening, a less compact remnant is found in the postmerger phase, the ejecta properties are altered, and a higher fraction of fast escaping material is observed. This material does not undergo leptonization, hence, we see that lighter r-process elements are underproduced, while DDLS$50$ and DDLS$70$ show a better agreement with respect to the solar abundance pattern. This might hint towards the possibility that electromagnetic counterparts could potentially distinguish the EOSs; however, for an accurate prediction, one would require more accurate opacities and heating rates, which is the reason we postpone this study to the future. In general, directly correlating this observable to $L_{\rm sym}$ in a conclusive way seems to be a non-trivial task, given the complex interplay between properties of the ejecta and features of the signal.

Our interpretation is based on the simulated tidal deformability $\Lambda_{\text {sim }}$, which reflects the thermal profile developed by the system during the simulation. This approach, rather than relying on the cold tidal deformability $\Lambda$, brings our results into agreement with theoretical expectations based on the stiffness of the EOS. Such a procedure proved valuable for the DDLS$30$ run, where thermal effects are prevalent, and with which we were able to consistently interpret our results. More precisely, we see that $\Lambda_{\rm sim}$ for DDLS$30$ is $\sim 17\%$ larger than its cold counterpart, closer to the stiffest DDLS$70$ configuration. The stiffening in our DDLS$30$ run can be seen in the similar merger times compared to DDLS$70$, and also in the post-merger dominant frequency $f_2$.

Next, we studied the imprint of $L_{\rm sym}$ on the GW signal. Comparing our numerical waveforms and waveforms produced with the \texttt{IMPhenomXAS\_NRTidalv3} approximant during the inspiral. we found good agreement, quantified via the dephasing between numerical and approximant signals, and showed that overall the dephasings are comparable throughout the simulations (within the error bands), with a better agreement for DDLS$30$ when using $\Lambda_{\rm sim}$, i.e., accounting for the thermal effects in the macroscopic properties of the NSs. Given that the approximant yields waveforms based on a few macroscopic binary parameters (mass, radius, and tidal Love number $k_2$ of the individual stars) and accurately reproduces the signal, we conclude that there is no clean imprint of $L_{\rm sym}$, besides the known role of $L_{\rm sym}$ in determining the NS structure~\cite{Li:2023bid}.

Considering the postmerger evolution, we find that the dominant $f_2$ peaks are very similar for DDLS$30$ and DDLS$70$, while for DDLS$50$, the difference of less than $\sim 50~{\rm Hz}$ with respect to the other setups suggests that distinguishing the EOSs would be difficult. In fact, previous studies point to measurement uncertainties of $50\sim100$ Hz for nearby BNS mergers with signal-to-noise $\mathcal{O}(10)$ during the post-merger~\cite{Rezzolla:2016nxn, Clark:2015zxa, Breschi:2022ens, Branchesi:2023mws}, i.e., distinguishing between these different scenarios will be challenging. 

We also compared our $f_2$ extracted from the simulations to the fitting formula [Eq.~(33)] of Ref.~\cite{Vretinaris:2019spn}, where $f_2$ is fitted against the chirp mass and effective tidal deformability. By doing so, we showed that (i) the fittings accurately reproduce our computed values (i.e., well within the fitting uncertainties), and (ii) that the overall agreement is improved with the use of $\Lambda_{\rm sim}$, in particular for DDLS$30$. Hence, this observation further strengthens the observation that the main contribution of $L_{\rm sym}$ to the GW properties enters via the tidal deformability, once again suggesting that distinguishing $L_{\rm sym}$ on the basis of GW signatures is unlikely. Moreover, our results point out that the description of DDLS$30$ in terms of $\Lambda_{\rm sim}$ is appropriate, although further work is necessary to understand if such a procedure is suitable for general EOSs with strong thermal features.\\

Finally, we want to put our results in comparison to the previous study of Ref.~\cite{Most:2021ktk}, where a systematic exploration of the impact of variations of $L_{\rm sym}$ in the evolution of BNS systems was conducted. For a clearer discussion, we briefly summarize the main methodological differences between the aforementioned work and our study. In the former, a generalized procedure for the construction of cold, catalyzed EOSs at high densities $n_b \geq 0.5~n_{\rm sat}$ in the form of piecewise polytropes is adopted. There, the authors treat the pressures at fiducial densities as free parameters for flexibility in the specification of $L_{\rm sym}$ and NS properties (e.g., radii and tidal deformabilities), and then proceed to the extension of their EOS to finite-temperature (keeping identical prescription for the thermal part) and arbitrary compositions as per Ref.~\cite{Raithel:2019gws}, while matching the low-density part of the EOS to SFHo EOS~\cite{Steiner:2012rk}. In doing so, their high-density EOSs are substantially different for their chosen $L_{\rm sym} = \{40, 100, 120\}~{\rm MeV}$ (see their Fig. 2), contrary to our EOSs, where both high- and low-density parts match for all models (see Fig.~\ref{fig:MRL-curves.pdf}). We restrict our comparisons to their $q = 1$, $R_{1.4} = 12~{\rm km}$ set of simulations, for which their total binary mass is $M = 2.71~M_\odot$, and the effective tidal deformabilities are $\tilde{\Lambda} = \Lambda =\{537, 395, 372\}$ from smallest to largest $L_{\rm sym}$. That is to say, their configurations are substantially softer than ours, and also, contrary to our case, their EOSs exhibit decreasing stiffness with increasing $L_{\rm sym}$. It is also important to notice that in our simulations, the NSs radii are different for the different EOSs; thus, direct comparisons are possible on a qualitative level, given that the dynamics of the remnant correlates to both tidal deformability and compactness.

We begin by noting that their reported differences for the post-merger waveforms (depicted in their Fig.~9) are more drastic than in our simulations. This is because the dynamics of the remnant, as argued by Ref.~\cite{Most:2021ktk}, seem to be strongly governed by the high-density EOS, which is identical in our setups. This seems to be the most relevant difference for the GW analysis, suggesting that our results are not in tension.

As a consequence of the identical high-density EOSs in our case, and the small spread of tidal deformabilities probed in our simulations, we do not find large differences in the post-merger peak $f_2$, but a fairly good agreement with respect to the fits of Ref.~\cite{Vretinaris:2019spn}. In contrast, Ref.~\cite{Most:2021ktk} identified a pronounced shift of $\sim 500~{\rm Hz}$ between their smallest and largest $L_{\rm sym}$ EOS.

Another important difference is found with respect to the properties of the ejecta. This is explained by observing that they model weak interactions via a leakage prescription~\cite{Ruffert:1995fs, Rosswog:2003rv}, while we employ a moment scheme~\cite{Shibata:2011kx, Foucart:2015gaa, Radice:2021jtw, Schianchi:2023uky}. This is a relevant distinction, as the thermal and compositional properties of both the remnant and the ejecta are significantly impacted by differences in neutrino treatment~\cite{Zappa:2022rpd}.

\section{Summary}

The key take-away messages from this work are summarized below. Firstly, the use of tidal deformabilities corrected for thermal effects appears to reconcile the well-established description of GW features with respect to binary parameters, as seen in the better agreement of the post-merger peak frequency $f_2$ and of the inspiral waveform when using $\Lambda_{\rm sim}$. Second, our results suggest that the effect of $L_{\rm sym}$ in the GW signal is almost completely (within the uncertainties of the simulations, fitting formulae, and waveform approximant) described by the modification of $\Lambda$. Thus, it seems unlikely that $L_{\rm sym}$ can be reliably distinguished by GWs. However, given the sufficiently distinct properties of the ejecta found in our simulations and the diverse abundances of r-process elements, we believe that the electromagnetic counterparts (e.g., kilonova) could allow us to identify $L_{\rm sym}$ if more accurate heating rates and opacities become available. Nevertheless, establishing a definitive correlation between specific signal features and the slope of symmetry energy remains a challenging task.

\section*{Acknowledgements}
HG and TD acknowledge funding from the EU Horizon under ERC Starting Grant, no. SMArt-101076369. Views and opinions expressed are those of the authors only and do not necessarily reflect those of the European Union or the European Research Council. Neither the European Union nor the granting authority can be held responsible for them.
The simulations with \texttt{BAM} were performed on the HPC system Emmy of
the North German Supercomputing Alliance (HLRN) [project nip00076], on the DFG-funded research cluster jarvis at the University of Potsdam (INST 336/173-1; project number: 502227537) and on the SDumont supercomputer of the National Laboratory for Scientiﬁc Computing (LNCC/MCTI, Brazil) [project 237565].
AS acknowledges the funding through the 
Polish National Science Centre (NCN) Grant 2023/51/B/ST9/02798
and the Deutsche For\-schungs\-gemeinschaft (DFG)
Grant No. SE 1836/6-1. 
\appendix

\end{document}